\documentclass{article}
\usepackage{graphicx,graphics}
\usepackage{url}

\begin{document}

\title{On the algorithmic behaviour of complex physical systems}
\author{Dan Anastasiu Pop\footnote{Pedagogical Highschool "Gh. Sincai", Zalau, Romania}, Gabriela Raluca Mocanu\footnote{Faculty of Physics, Babes-Bolyai University, No. 1
Kog\u{a}lniceanu Street,400084, Cluj-Napoca, Romania,
gabriela.mocanu@ubbcluj.ro}, George Arghir\footnote{Department of Materials Science and Technology, Technical University, Cluj Napoca}}

\date{}

\maketitle

\begin{abstract}
Starting from the idea that the underlying mechanisms driving the observable processes in nature is algorithmic, we exemplify this in two ways: nature works as a computing machine and thus the processes running on it optimize themselves in an intrinsic manner so as to save time. As a first example we will place the empirical analysis of oxidation in the context of algorithm complexity analysis. Second, we will show that deformations suffered by a shape-memory alloy may be produced by nature showing a LIFO (Last In First Out) behaviour. 
\end{abstract}

\section{Introduction\label{intro}}

The role of computers in simulations of natural processes has been growing, but from our point of view it is crucial to separate between two paradigms. The first paradigm is best described as: the model of the physical process is known and consists of evolution rules and some machine is used to numerically implement the model. The second differs from the first in a subtle way: the natural processes behave as if they were a machine and this is the framework we adhere to.

Information has become a conceptual tool above others and it is found everywhere across
research disciplines and in everyday use~\cite{ref:1}. According to pancomputationalists the dynamics of the universe is a computational process; universe on the fundamental level may be conceived of as a computer which from the current state, following physical laws, computes its own next state~\cite{ref:1}. This is called "natural computing" and we see this as an extension of "physics" to "physical phenomena are computational processes". 

We will explore two straightforward examples of the above concepts, coming from the field of material science. We focus on apparently simple physical phenomena as it is in these types of manifestations that the intrinsic algorithmic becomes evident. More complex phenomena will appear as result of a superposition of simple algorithms. They may be of course investigated in the framework we are proposing, but such an investigation is at this point beyond the purpose of our work. 

The first phenomenon we will discuss is the oxidation of metallic surface. Oxidation appears as a consequence of Nature minimizing its algorithm complexity. In the same spirit, alloys show memory when they are deformed as a result of the data processing mechanism.

The paper is organized as follows: in Section~\ref{sec2} we present the phenomenological concept of oxidation and the theoretical concepts of algorithm complexity. In Section~\ref{sec3} we present the processes suffered by a shape memory alloy and the data access concept of Last In First Out. The conclusions of this work follow in Section~\ref{sec4}.

\section{Oxidation and algorithm complexity\label{sec2}}

As a starting point we will use oxidation rates of pure metals. There have been three basic kinetic laws proposed to describe the oxide film thickness growth as a function of time~\cite{ref:2}. These are the parabolic rate law, derived from Fick's first law of diffusion and two empirical laws: the logarithmic rate law and the linear rate law. We will write these laws equating the time out ($t$), and not the film thickness/mass gain ($x$) as is usually done in the literature. 

The parabolic rate law states that

\begin{equation}
t = \frac{x^2-x_0}{k_p},\label{eq:1}
\end{equation}
i.e., $t$ behaves as $x^2$; the logarithmic rate law states that
\begin{equation}
t = \frac{1}{c}\exp \left \{ \frac{x}{k_p} \right \} - \frac{b}{c},
\end{equation}
i.e., $t$ behaves as $\exp {x}$, while the linear rate law states that time grows linearly with mass gain
\begin{equation}
r = \frac{x}{k_L},\label{eq:3}
\end{equation}
where $x_0$, $k_p$, $b$, $c$, $k_L$ are constants.

\begin{figure}[!h]
\includegraphics[scale=1]{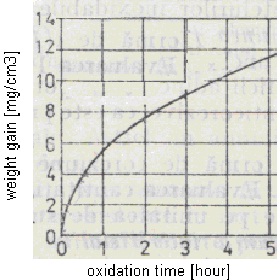}
\caption{Time variation of the mass gain of Copper and the influence of different alloying elements~\cite{ref:3}.}\label{fig:fig1}
\end{figure}

As an illustration for the temporal evolution of mass gain in Copper undergoing oxidation in the case of the parabolic growth rate, see Figure~\ref{fig:fig1}.

We propose that nature is algorithmic in nature and it is optimized to run the process so as to minimize time, i.e., the amount of time required to process the data is minimal under the given initial and boundary conditions. In computer science, algorithm complexity gives the time necessary to run an algorithm as a function of the number of times a basic instruction is executed. Such calculations are an important tool to discriminate between the temporal efficiency of conceptually different algorithms which solve the same problem. 

We thus proceed to state that if the empirical behaviour of a process looks like the one in Figure 1, than the algorithm “running” in the background to lead to the observed process has complexity $\mathcal{O}\left( x^2 \right)$, meaning that the time needed to run a programme behaves as $n^2$. In this context we report general well known results of algorithm complexity theory~\cite{ref:3}, namely that commonly used computing times are, corresponding to Equation~\ref{eq:1}, $\mathcal{O}\left( n^2 \right)$, corresponding to Equation~\ref{eq:3}, $\mathcal{O}\left( n \right)$, and others such as $\mathcal{O}\left( n^3 \right)$, $\mathcal{O}\left( 2^n \right)$.

\section{Shape-memory alloys and LIFO\label{sec3}}

\begin{figure}[!h]
\includegraphics[scale=0.7]{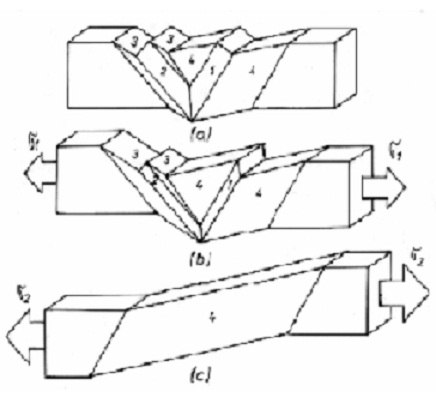}
\caption{De-twinning mechanism in a grain with a multivariate martensitic structure from a  polycrystalline shape memory alloys (PMA): (a) group of self-accommodated martensite plates heat induced, (b) partial de-twinning
produced by increased tensile stress, $\sigma _1$, (c) complete de-twinning~\cite{ref:3}.}\label{fig:2}
\end{figure}

The physics of shape-memory alloys deformation can be summarized as follows~\cite{ref:3}: look at Figure~\ref{fig:2}. 

Consider a grain with a multivariate martensitic structure, from a  polycrystalline shape memory alloy, subjected to traction. Following cooling, a group of four variants of auto-accommodating thermically induced martensite is considered to have been formed. Variants are accommodated by twinning, which means that they mutually adjust their volume to be entered in the space available in the austenitic matrix (much tougher and more rigid). 

A crystalline grain may present with up to six different groups with various orientations, giving a maximum of 24 versions. In Figure~\ref{fig:2}(a) it can be seen that deformations caused by variations in the pairs' being in "twinning relationship" are equal and opposite. Basically, the 1-4 pair formation automatically leads to the 2-3 pair formation. Thus the variation of macroscopic total volume is zero. When applying a tensile stress (at $T =$ constant) the most favourable oriented to the axis of tension of
martensite variants is developed, in relation to Schmid's law. In Figure~\ref{fig:2}(b) these variants were taken to be 3 and 4. It is to be noted that their development is at the expense of other variants, 1 and 2, which practically disappear. So the application of tensile stress leads to a partial de-twinning group variants of martensite plates. If the tensile stress is increased, the material elongates, while full de-twinning group occurs leading to one version, 4, which has the most favourable orientation. Continued application of the stress leads, in a first step, in elastic deformation of the most favourable oriented version. In this situation, the result of a decrease of the tensile stress, if no sliding occurs in the process, is that the same phenomena occur in reverse order~\cite{ref:5,ref:6}.

\begin{figure}[!h]
\includegraphics[scale=1]{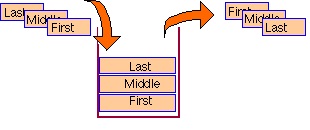}
\caption{Schematic representation of how data is accessed in a LIFO-based algorithm (Figure from~\cite{ref:8}).}\label{fig:3}
\end{figure}

In computer science, the concept “Last In First Out” (LIFO) refers to the way items stored in some types of data structures are processed~\cite{ref:7}. The abstraction LIFO (Figure~\ref{fig:3}) represents for our purposes a set of rules of list processing and temporary storage, which is done \emph{in a certain order}. The ability to organize data by order rearrangement is in itself a great tool to assure the efficiency and flexibility of the Computing machine.

We thus state that the behaviour of shape-memory alloys is just a consequence of how the “Computing” space processes data in the case of those materials. This may in fact be easily seen if you look at Figure~\ref{fig:2} again having the benefit of this new hindsight.

\section{Conclusions\label{sec4}}

We argued that the observable processes of oxidation and shape-memory alloys have as an underlying cause the algorithmic behaviour of Nature seen as a Computing Machine. Oxidation exhibits specific laws of mass growth to as to minimize computing time according to algorithm complexity calculations, thus explaining growth laws that were up to now considered empirical. Shape-memory alloys appear as a consequence of how the “Computing” space processes data in the case of those materials.

Future work in this area should be directed identifying fundamental (perhaps yet unknown) physical concepts based on the (confirmed) existence of algorithmic concepts. One of these future research ideas may be the link between recorded temporal duration of a process and the number of elementary operations this process would need to run on a computing machine (time is intrinsically quantified in units of the simplest operation running on a machine).

\end{document}